\documentclass[useAMS,usenatbib]{mn2e}
\usepackage{amssymb,amsmath,graphicx}
\usepackage{longtable}[1]
\usepackage{graphics}
\usepackage{threeparttable}
\usepackage{lscape}
\usepackage{color}

\title[The old globular cluster system of NGC\,4365: new VLT/FORS2 spectra]{The old globular cluster system of NGC\,4365: new VLT/FORS2 spectra}
\author[A. L. Chies-Santos, S. S. Larsen and M. Kissler-Patig]{A. L. Chies-Santos$^{1}$\thanks{E-mail:
Ana.Chies\_Santos@nottingham.ac.uk}, S. S. Larsen$^{2}$ and M. Kissler-Patig$^{3,4}$\\
$^{1}$University of Nottingham, School of Physics and Astronomy, University Park, NG7 2RD Nottingham, UK\\
$^{2}$Department of Astrophysics/IMAPP, Radboud University Nijmegen, PO Box 9010, 6500GL Nijmegen, the Netherlands\\
$^{3}$European Southern Observatory, 85749 Garching bei M\"unchen, Germany\\
$^{4}$Gemini Observatory, 670 N. A'Ohoku Place, Hilo, HI 96720, USA
}
\begin{document}

\date{Accepted 2012 September 13; Received 2012 September 7; in original form 2012 August 17}

\pagerange{\pageref{firstpage}--\pageref{lastpage}} \pubyear{2011}

\maketitle

\label{firstpage}

\begin{abstract}
We present new spectroscopic observations of a sample of globular clusters (GCs) in the Virgo Cluster elliptical NGC\,4365. The great majority of the objects analysed here, overlap on purpose with objects presented in two previous studies that came to contradictory conclusions.  
We measure Lick indices for the GCs and infer ages, abundances and $\alpha$-enhancement values by comparing to population synthesis models. 
Our results do not support the existence of numerous intermediate-age GCs in this galaxy as suggested by some of the past studies. Instead, we find ages, consistent with the range $\sim 8-14$\,Gyrs (model dependent and with large uncertainties).
We also find that the newly observed GCs have [Fe/H]$\sim$$-1.3$ to $0.3$ dex and the majority are consistent with [$\alpha$/Fe]$\,\sim\,0.3$ dex, independent of metallicity.
By comparing the metal sensitive index [MgFe] with optical colours, we find that our data is consistent with 14\,Gyr models. However for optical/near-infrared colours we find that the same models would have to be redder by $\sim0.2$\,mag in order to be consistent with our data. This offset is not found to be consistent with an age effect  as it is even larger for younger models. 
\end{abstract}


\section{Introduction}
Extragalactic globular clusters (GCs) are often employed as tracers of star formation histories of galaxies (see e.g. \citealt{harris91}; \citealt{kp97}; \citealt{gebhardt99}; \citealt{larsen01}; \citealt{bs06}).
They represent discrete star formation probes and as such are more easily modeled than the mix of stellar populations that make up the integrated light of their host galaxies. 
Through long exposures in multi-slit spectrographs at 8-10\,m telescopes it is possible to obtain high quality spectra of GCs to measure line strength indices and through these infer their ages, metallicities and chemical abundances.
Knowledge of such properties of GCs is of utmost importance in order to constrain co-formation theories of GCs and galaxies.
However, the task of using extragalactic GC spectroscopy for stellar population purposes is challenging. Except for the Milky Way (\citealt{puzia02b}; \citealt{schiavon05}); M\,31 (eg. \citealt{strader11}) and NGC\,5128 (\citealt{woodley10}) it is very difficult to obtain large samples of high-quality integrated spectra and overcome certain selection biases (\citealt{puzia03}; \citealt{strader05}; \citealt{bs06} and references therein). Generally, for galaxies at Virgo and Fornax Cluster distances, only the brightest GCs can be measured. Moreover, the GCs in the very central parts have to be avoided because of the strong galaxy background light. Therefore, extragalactic GC studies have to rely on photometric studies combined with generally small and most likely non-representative (of the GC system as a whole) spectroscopic samples. 

The giant elliptical NGC\,4365 represents a puzzling case in the literature of the field of extragalactic GCs.
It became an interesting target due to the somewhat unusual colour distribution of its GC system (\citealt{dforbes96}; \citealt{larsen01}).
In general, the GC systems of giant elliptical galaxies exhibit bimodal optical colour distributions, where the colours of both peaks show a fairly tight correlation with host galaxy
integrated luminosity (\citealt{larsen01}; \citealt{peng06}). NGC 4365 has been reported to have a normal blue peak for its luminosity, but a broader read peak shifted towards bluer $(V-I)$ colours (\citealt{larsen01} with WFPC2 data, \citealt{lbs05} with ACS imaging). \cite{peng06} report bimodality for the $(g-z)$ colour distribution of this galaxy with a broad red peak. The clusters that turn the red peak somewhat bluer are very close to the centre (\citealt{lbs05}). \cite{blom12} presents a wide field imaging optical survey of the GC system of NGC\,4365 and find support for three distinct GC colour populations.
From $BIK$ photometry, \cite{puzia02} suggest the presence of a great fraction of intermediate-age GCs in this galaxy. This issue has also been investigated with optical/near-infrared photometry by \cite{lbs05} and \cite{kundu05}. While the latter strongly favours the presence of intermediate-age GCs, the former presents evidence both for and against intermediate-ages.

Spectroscopic observations of GCs in this galaxy have so far failed to provide a conclusive answer with one study suggesting intermediate ages (\citealt{larsen03}, hereafter L03) and another yielding uniformly old ages (\citealt{brodie05}, hereafter B05). These studies analysed two sets of observations of a similar sample of clusters with the LRIS spectrograph on the Keck telescope. The L03 sample consisted of 14 GCs and the B05 one of 22 GCs. The S/N of B05 was higher than that of L03. We refer the reader to Sect.\,4.8 of \cite{lbs05} for a thorough discussion of a comparison between the L03 and B05 samples and analysis.     
Integrated light spectroscopic studies show that the galaxy light is dominated by old and metal-rich stars as in most other giant ellipticals (\citealt{davies01}; \citealt{yamada06}). Therefore a large fraction of younger GCs would be surprising.
Most giant ellipticals are thought to have at place the bulk of their GC systems at $z=2$ (\citealt{puzia05}; \citealt{strader05}). The claims of numerous intermediate-age ($3-5$\,Gyrs) GCs in NGC\,4365 would imply that a major star formation episode occurred at a lower redshift ($z<2$) in this galaxy.
It is therefore, of critical importance to verify such claims.
A large number of intermediate-age GCs without a corresponding intermediate-age (field) stellar population would imply that GCs can form in large numbers even in a modest star formation event.
More recently however, \cite{paper2} based on optical/near-infrared photometry showed through an empirical approach that the age distribution of the GC system of NGC\,4365 is no different from that of other large ellipticals (e.g. NGC\,4486 and NGC\,4649).

In this paper a third spectroscopic analysis of GCs in NGC\,4365 is presented with the purpose of shedding light on the contradicting claims on their ages. 
Metallicities and chemical compositions of this new sample of GCs are also studied.
While we provide new observations, most of the GCs analysed here overlap with those of L03 and/or B05, as we intend to understand why these two studies came to different conclusions.

\section{Observations and data reduction}
Spectra for GCs in NGC\,4365 were obtained using the mask exchange unit (MXU) of the Focal Reducer and Low Dispersion Spectrograph (FORS2) on the Very Large Telescope (VLT) UT1. 
The observations were carried out in service mode during several nights of Period 78A (2007) spread between February and March.
The observed 22 GC candidates were selected to overlap the Keck/LRIS studies of L03 and B05.
All exposures used the 600B+22 grism with 600 grooves per mm and a slit of 1$\arcsec$. 
The wavelength coverage of the system is 3400\AA - 6100\AA. 
The dispersion is of 0.66\,\AA/pix, pixel scale $0.126\arcsec$pix and the final resolution is $\sim5$\,\AA.

The calibration data for each night were ran through the pipeline recipe $fors\_calib$ in ESOREX. 
With the recipe $fors\_science$ the scientific spectra were reduced applying the extraction mask and the normalised flat-field created with $fors\_calib$.
Heliocentric velocity correction was performed for each extracted spectrum of every exposure.
The individual spectra of each GC were finally co-added, resulting in a total exposure time of 548.5 minutes ($\sim$\,9 hours).
Using the FXCOR task in the RV package in IRAF, radial velocities were determined by cross-correlating the GC spectra with a model spectrum. 
The GC spectra were shifted to zero radial velocity.
Similarly, the reduction of 6 Lick standard stars and 6 radial velocity standards, observed in long slit mode was performed with $fors\_calib$ and $fors\_science$.
Finally, flux calibration was applied to the Lick standard stars and GC spectra through the tasks \textit{standard}, \textit{calibrate} and \textit{sensfunc} within the ONEDSPEC IRAF package.

\begin{figure}
\begin{center}
\includegraphics[width=8cm]{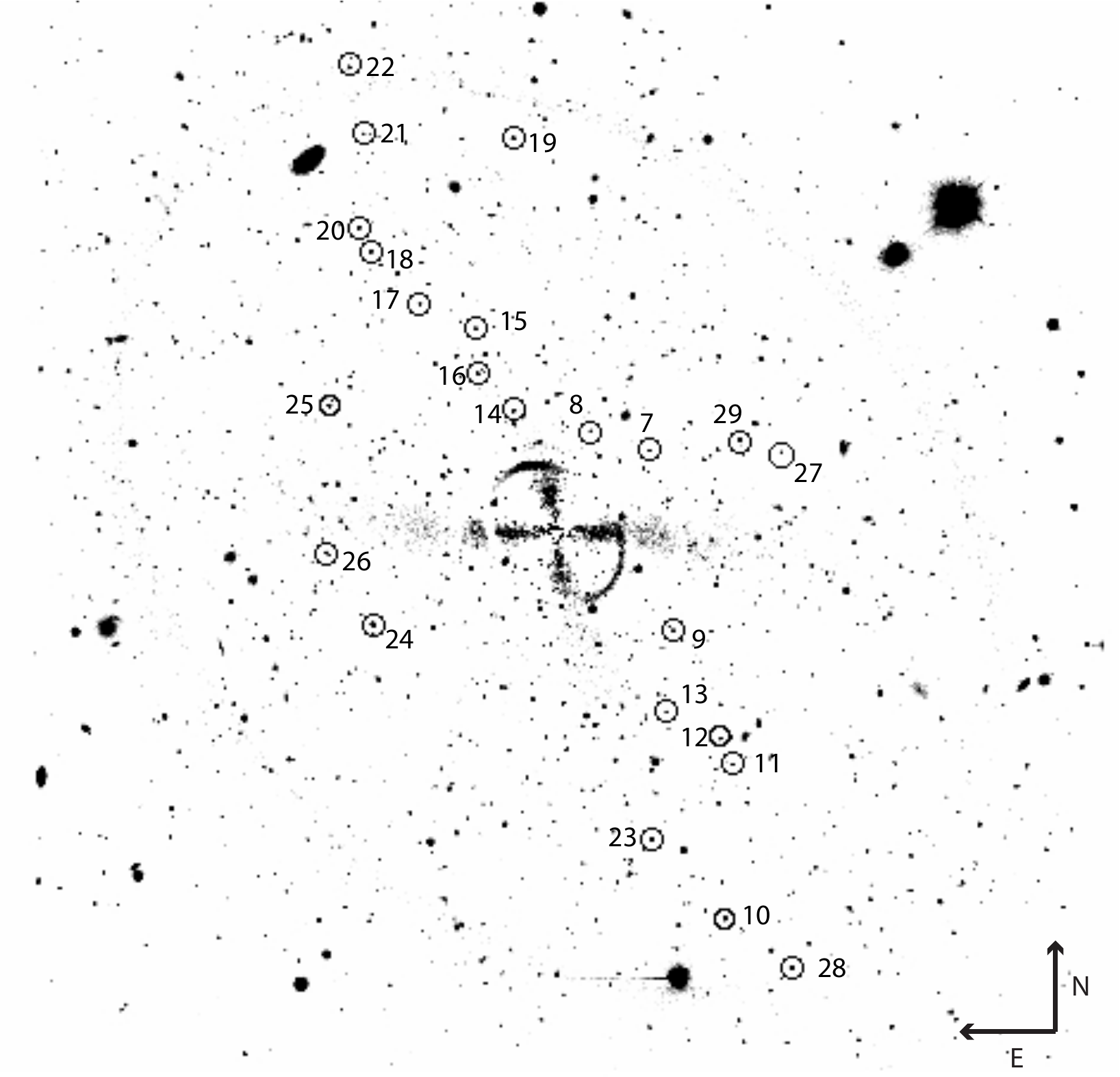}
\caption{A model subtracted FORS2 image of NGC\,4365 with the observed GC candidates indicated as circles and by their corresponding IDs from Table \ref{GCs}. The image size is the field of view of FORS2: $6.8\arcmin\,\times\,6.8\arcmin$. The cross-like feature in the centre of the image is a residual from the model subtraction.}
\label{image}
\end{center}
\end{figure}

\begin{figure}
\begin{center}
\includegraphics[width=9cm]{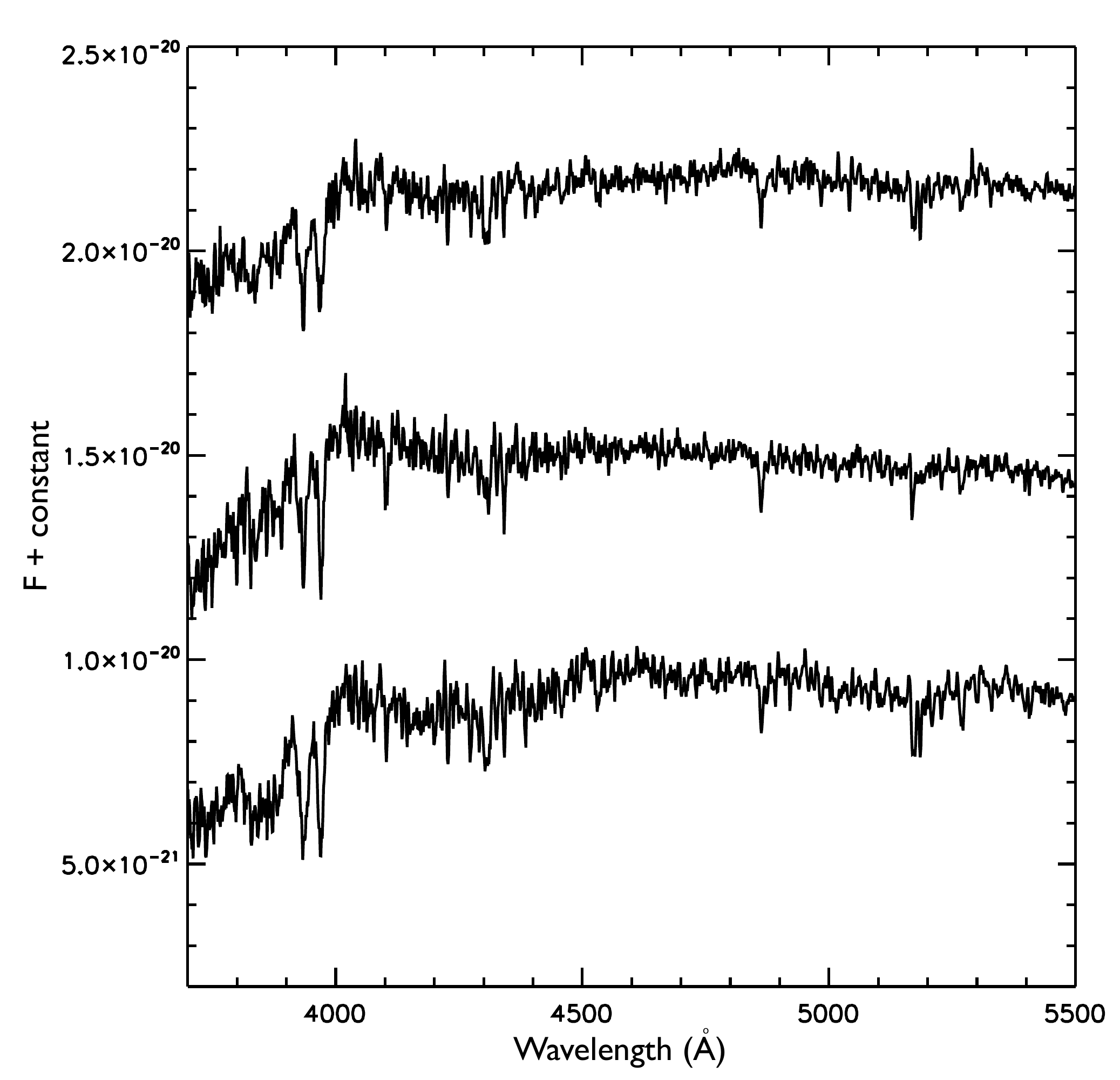}
\caption{Sample spectra of clusters of average S/N (from top to bottom: ids 11, 25 and 17), shifted to zero radial velocity.}
\label{samplespectra}
\end{center}
\end{figure}

In Fig.\,\ref{image} the observed GC candidates are indicated in a model subtracted FORS2 image of NGC\,4365. The model image was created with the ELLIPSE and BMODEL tasks in the STSDAS package in IRAF.
In Table\,\ref{GCs} we list the observed globular clusters, their equatorial co-ordinates, the correspondence to L03 and B05, their radial velocities and the signal to noise ratio (S/N) of the final spectra.
Based on the derived radial velocities, objects 26 and 29 are obviously non-clusters. These are removed from further analysis.
Fig.\,\ref{samplespectra} shows three sample spectra with average S/N.

\begin{table}
\begin{scriptsize}
\begin{center}
\label{GCs}
 \caption{Globular cluster candidates observed with FORS2: (1): id, (2) and (3): RA, Dec, (3): id from L03 and/or B05, (5): radial velocity and (6): S/N per pixel at $4500\,\lesssim$\,$\lambda$(\AA)\,$\lesssim\,5000$.}
 \label{GCs}
 \begin{tabular}{cccccc}
  \hline
   (1) & (2)& (3) & (4) &(5) & (6) \\
  \hline
 ID & RA (J2000) & Dec (J2000) & B05 \& L03& $V_r$(km/s)&S/N\\
  \hline
07  &    12:24:25.89   &    7:19:36.68&  L03-09         &  765   $\pm$   8    &  33.1   \\ 
08  &    12:24:27.41   &    7:19:44.23 & L03-11, B05-13 &  476	$\pm$  12  &  20.6  \\ 
09  &    12:24:25.29   &    7:18:27.52 & L03-05, B05-08 &  1580   $\pm$ 33 & 18.8 \\
10  &    12:24:23.98   &    7:16:37.46 & B05-1          &  723	$\pm$   16 &  51.2   \\ 
11  &    12:24:23.76   &    7:17:36.53 & B05-3          &  601	$\pm$  7   &  33.2   \\ 
12  &    12:24:24.07   &    7:17:46.38 & L03-01         &  1212  $\pm$  12  &  26.6  \\ 
13  &    12:24:25.46   &    7:17:56.62 & L03-02, B05-06 &  1344  $\pm$   23 &  19.2  \\ 
14  &    12:24:29.40   &    7:19:52.10 & L03-13         & 1044  $\pm$  16  &  28.0 \\ 
15  &    12:24:30.36   &    7:20:23.42 & B05-17         & 868	$\pm$  21  &  24.9    \\ 
16  &    12:24:30.31   &    7:20:06.34 & L03-15         & 865	$\pm$   19 &  32.6   \\ 
17  &    12:24:31.80   &    7:20:32.90 & B05-18         & 1176  $\pm$  12  &  35.2  \\ 
18  &    12:24:33.07   &    7:20:53.23 & B05-19         & 1213  $\pm$  17  &  34.1  \\ 
19  &    12:24:29.40   &    7:21:36.48 & L03-16         & 1113  $\pm$  14  &  44.8  \\ 
20  &    12:24:33.36   &    7:21:02.23 & B05-20         & 1443  $\pm$  12  &  39.6  \\ 
21  &    12:24:33.24   &    7:21:38.28 & B05-22         & 903	$\pm$   15 &  20.4    \\ 
22  &    12:24:33.60   &    7:22:03.61 & B05-23         & 999  $\pm$   55 & 19.5 \\ 
23  &    12:24:25.85   &    7:17:07.63 & -              & 974	$\pm$  17  &  36.3    \\      
24  &    12:24:30.77   &    7:17:47.68 & -              &   910   $\pm$  22  &  26.6  \\ 
25  &    12:24:34.13   &    7:19:54.11 & -              &  866	$\pm$  39  &  32.5    \\ 
26  &    12:24:33.03   &    7:18:30.13 & -              &   -6	    $\pm$   10 &  83.2   \\ 
27  &    12:24:22.51   &    7:19:36.04 & -              &  732	$\pm$   98 &  23.5   \\
28  &    12:24:22.24   &    7:16:18.74 & -              &  1412  $\pm$  9   &  51.5  \\ 
29  &    12:24:23.57   &    7:19:40.57 & -              &  119   $\pm$   9  & 55.5          \\ 
 \hline
 \end{tabular}
  \end{center}
\end{scriptsize}
 \end{table}

To measure the Lick/IDS indices of the GCs we employ the Lick\_EW code of \cite{graves08}.
It measures the equivalent widths of absorption indices from \cite{worthey94a}, including smoothing the spectra to the Lick/IDS resolution. 
 
\begin{figure*}
\begin{center}
\includegraphics[width=9cm, angle=90]{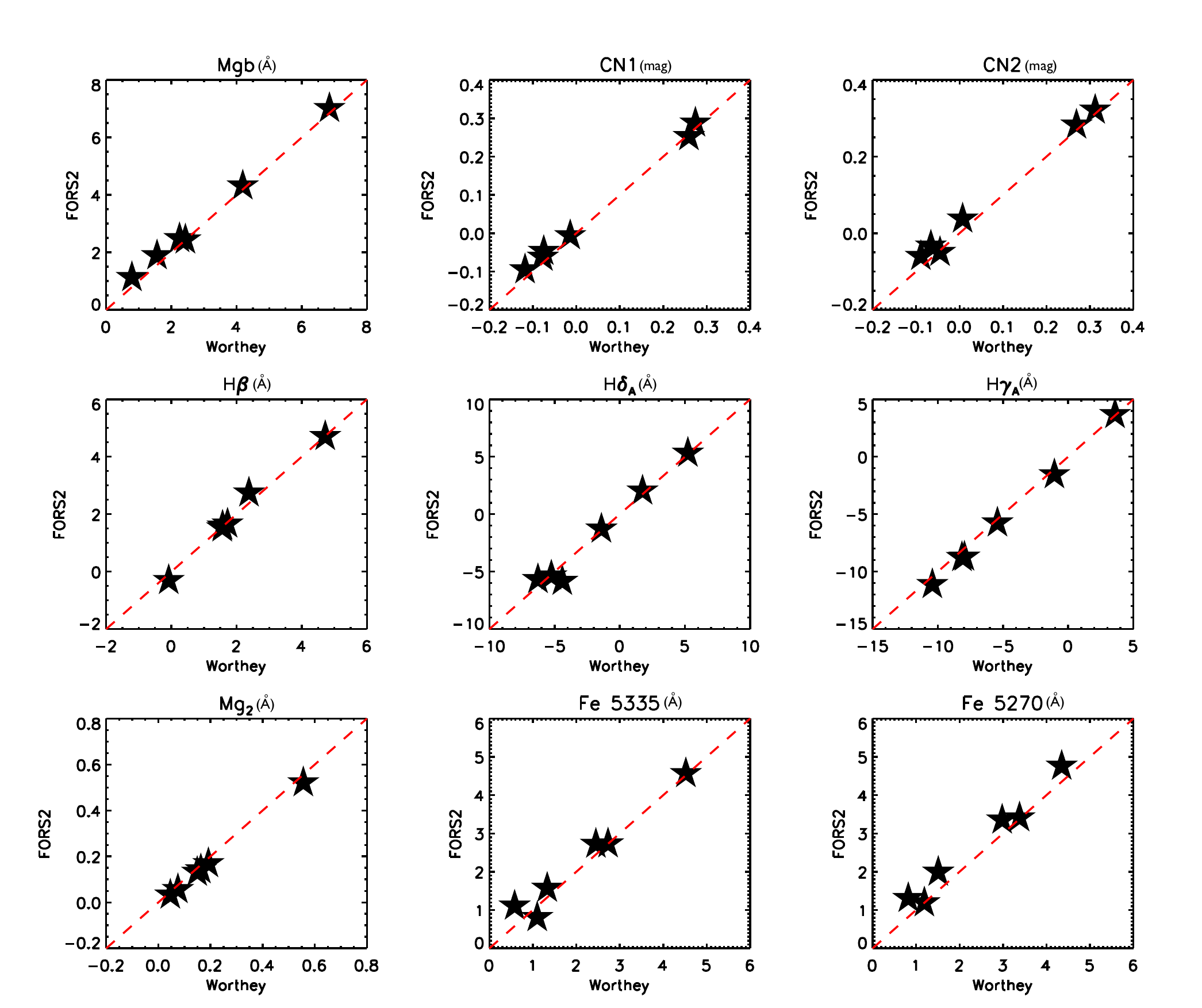}
\caption{Comparison of some of our Lick/IDS index measurements for six Lick standard stars and the \protect \cite{worthey94a}. The dashed lines are one-to-one relations.}
\label{lickstds_compare}
\end{center}
\end{figure*}

Fig.\,\ref{lickstds_compare} shows the comparison between some of our Lick/IDS measurements and those of \cite{worthey94a} for six Lick standard stars.
The offsets between the two systems are very small and compatible with measurement errors alone. Thus we do not apply the offsets to the spectra. They are shown in Table\,\ref{lickstds_compare_tab}.

\begin{table}
\begin{scriptsize}
\begin{center}
 \caption{Offsets between our Lick standard stars and the values from \protect \cite{worthey94a}.}
 \label{lickstds_compare_tab}
 \begin{tabular}{cc}
  \hline
 Index & $<$measured-standard$>$ (\AA/mag)\\
  \hline
Mgb &    0.201 $\pm$ 0.052 \\
CN1&   0.014 $\pm$  0.005\\
CN2 &   0.019  $\pm$ 0.006\\
H$\beta$ &   0.011 $\pm$ 0.080\\
H$\delta_{A}$ &   -0.067   $\pm$  0.287\\
H$\gamma{A}$ &   -0.459  $\pm$ 0.129\\
Mg$_2$ &   -0.021  $\pm$ 0.003\\
Fe 5335 &    0.134   $\pm$ 0.114\\
Fe 5270 &    0.299   $\pm$ 0.009\\
 \hline
 \end{tabular}
 \end{center}
 \end{scriptsize}
 \end{table}

\section{Results}
In this section we compare the Lick indices of the GCs with simple stellar population (SSP) model grids.
Several models exist in the literature, here we use the state-of-the-art models \cite{schiavon07} (from now on referred to as S07), and \cite{tmj11} (from now on refered to as T11). 
There are five S07 metallicity values ([Fe/H]=-1.3 -0.7, -0.4, 0.0, 0.2) while there are six T11 ones ([Fe/H]=-2.25,-1.35 -0.33,0.0, 0.35, 0.67). Both models are available for different [$\alpha$/Fe] values and ages relevant for this study (in the range of 3 to 14\,Gyrs).

\subsection{Comparison with Keck/LRIS studies}
\begin{figure*}
\begin{center}
\includegraphics[width=13cm]{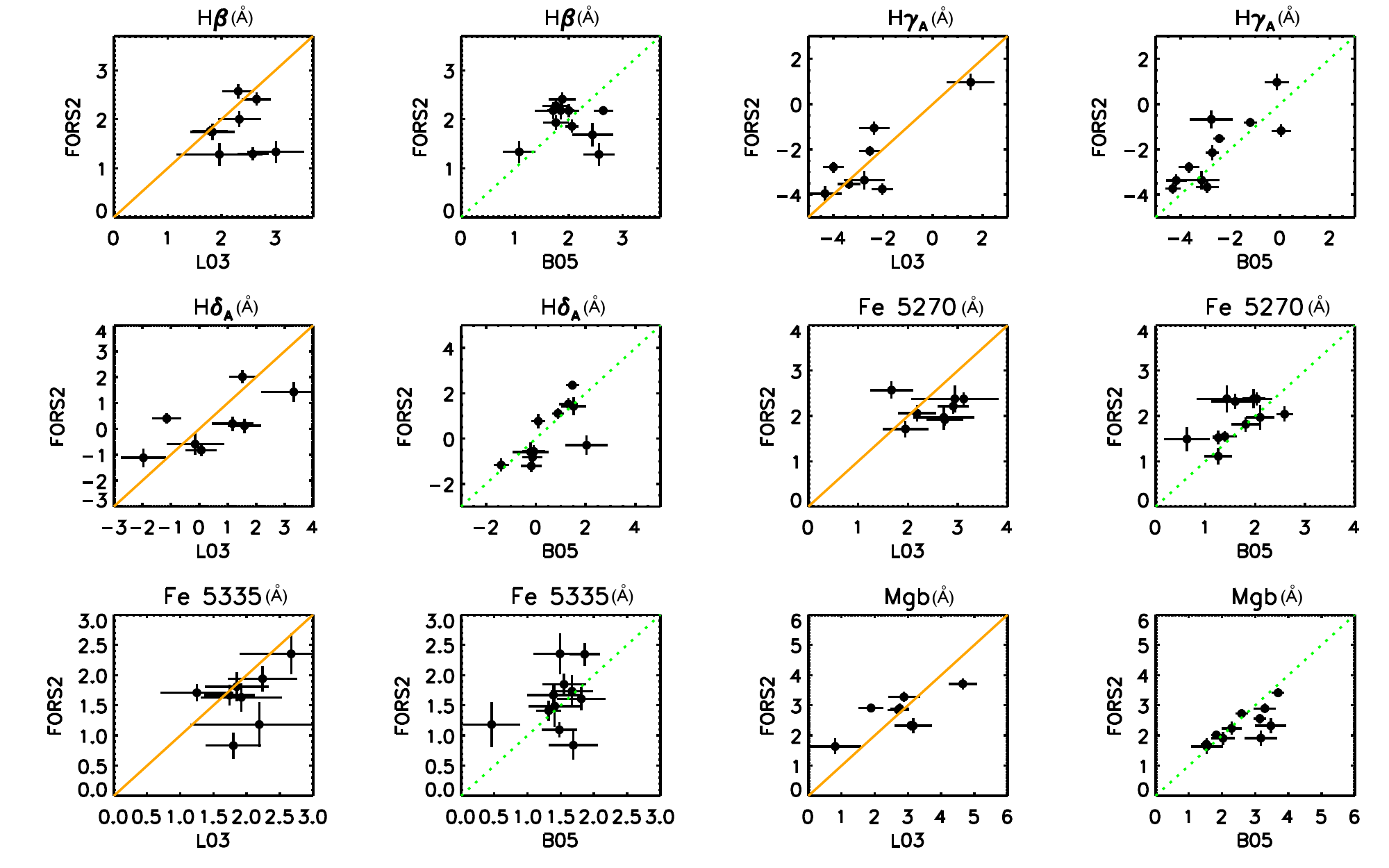}
\caption{Several L03 and B05 Lick/IDS index measurements \textit{vs.} the FORS2 ones for the GCs in common. The orange solid and green dotted lines indicate one-to-one relations between L03 and this study and between B05 and this study respectively.}
\label{FORS2_LRIS}
\end{center}
\end{figure*}

Fig.\,\ref{FORS2_LRIS} shows several L03 (and B05) Lick/IDS index measurements versus the FORS2 measurements for the GCs in common between L03 and FORS2 (and B05 and FORS2). There are 8 objects in common with the FORS2 and L03 sample while there are 11 in common with the B05 sample.
The lines in Fig.\,\ref{FORS2_LRIS} are one-to-one relations.
Overall there is good agreement between FORS2 and L03/B05 considering the measurement errors, especially for H$\gamma$, H$\delta_A$ and $Mgb$.
The mean and median differences between FORS2 and L03 and between FORS2 and B05 are shown in Table\,\ref{GCs_compare_tab}, where the errors are the standard error of the mean. 
It is clear from this table and from Fig.\,\ref{FORS2_LRIS} that the scatter is different for different indices. However, the scatter is smaller for the comparison between B05 and FORS2. The FORS2 measurements are closer to B05 than they are to L03.

\begin{table*}
\begin{scriptsize}
\begin{center}
\caption{The mean and median differences between L03 and FORS2 and B05 and FORS2. The errors are the standard error of the mean.}
\label{GCs_compare_tab}
 \begin{tabular}{ccccccc}
 \hline
 Index & median(L03$-$FORS2) (\AA) & mean(L03$-$FORS2) (\AA) & median(B05$-$FORS2) (\AA)& mean(B05$-$FORS2) (\AA)\\
  \hline
	H$\beta$     &  0.24 &     0.42  $\pm$    0.24 &-0.18 & 0.02  $\pm$ 0.18\\
	H$\gamma{A}$ &  0.17 &  -0.03   $\pm$    0.36 &-0.58 & -0.46  $\pm$ 0.27\\
	H$\delta_{A}$&  0.89 &     0.35 $\pm$     0.42 &0.09 &   0.25$\pm$  0.27\\
	Fe 5270      &  0.69 &     0.38 $\pm$     0.20 & 0.26 &  0.26 $\pm$  0.14\\
	Fe 5335      &  0.30 &     0.32 $\pm$     0.17 &0.09&   0.13$\pm$  0.15\\
	Mgb          & -0.14 &  0.0008  $\pm$    0.27 & 0.12 &   0.29 $\pm$ 0.15\\
\hline
\end{tabular}
\end{center}
\end{scriptsize}
\end{table*}

\subsubsection{Balmer line indices: inferring ages \& metallicities}
\label{balmer}

\begin{figure*}
\begin{center}
\includegraphics[width=18cm]{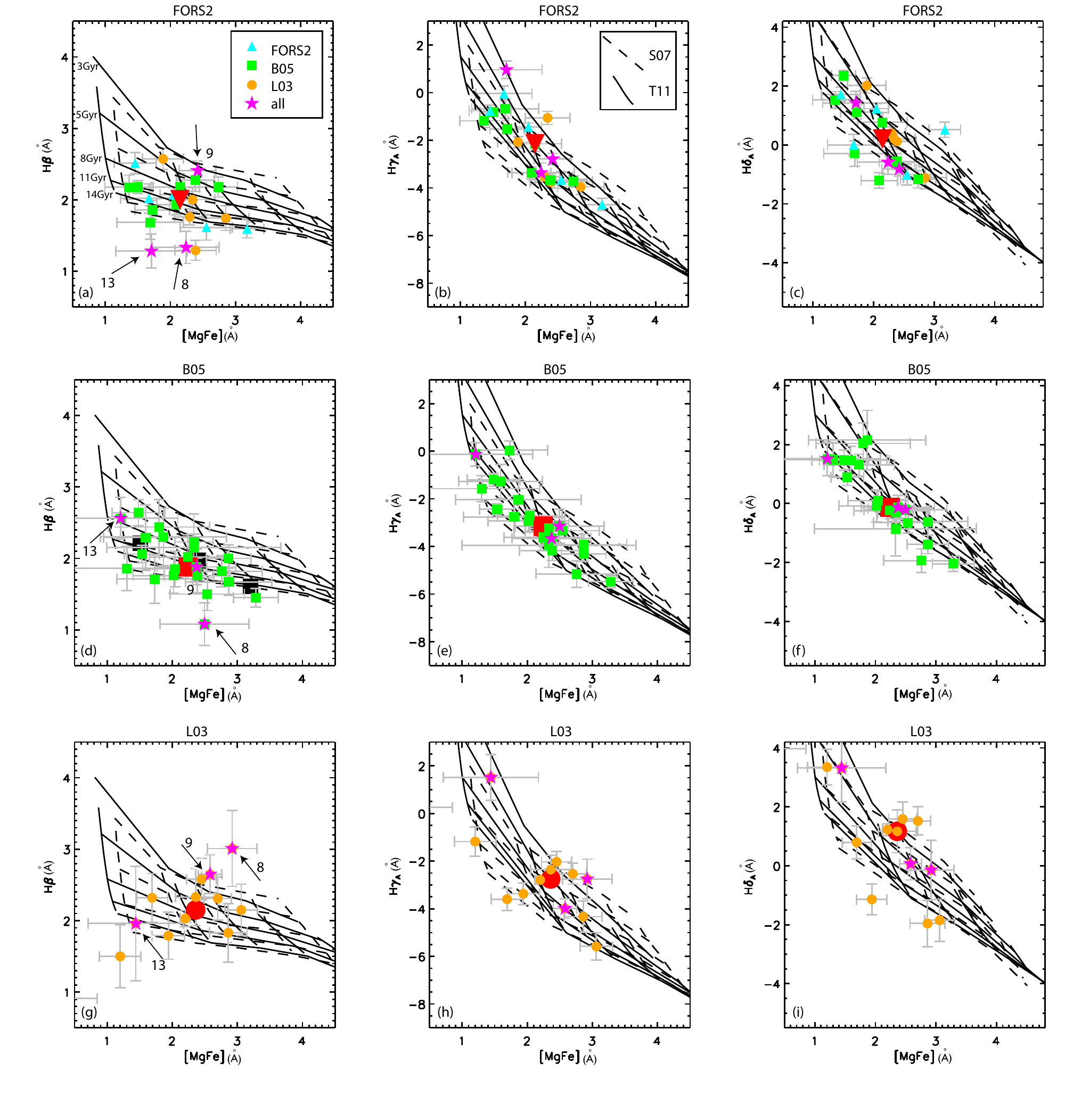}
\caption{[MgFe] \textit{vs.} $H_{\beta_A}$, $H_{\gamma_A}$ and $H_{\delta_A}$ diagnostic diagrams for GCs in NGC\,4365. Overlaid are [$\alpha$/Fe]=0.3 model grids from S07 (dashed lines, left to right [Fe/H]=-1.3 -0.7, -0.4, 0.0, 0.2) and T11 (solid lines, left to right, [Fe/H]=-1.35 -0.33,0.0, 0.35, 0.67). The ages are according to the legend in panel (a). 
\textit{upper panels:} the FORS2 sample with different colours and symbols indicating the GCs in common with the LRIS samples of L03, B05 and the new ones obtained with FORS2 according to the legend in panel (a). The upside-down red triangle represents the median value for the GCs in the different plots.
\textit{middle panels:} the B05 sample (represented by green squares). The black squares in panel (d) represent the medians of the B05 defined three populations, see text for details. The red square is the B05 median. 
\textit{bottom panels:} the L03 sample (represented by orange circles) where the red circle is the L03 median.
The GCs in common to the three samples have their ids indicated in panels (a), (d) and (g) according to the designation of Table\,\ref{GCs}.}
\label{indices_plot}
\end{center}
\end{figure*}

While individual absolute cluster ages and metallicities cannot be determined accurately as these are strictly model dependent, 
the comparison of Balmer line indices with metal sensitive ones can give an estimate for ages and metallicities.
In Fig.\,\ref{indices_plot} diagnostic diagrams [MgFe] \textit{vs.} Balmer line indices (H$\beta$, H$\gamma_{A}$ and H$\delta_{A}$) are shown for GCs in NGC\,4365.
Overlaid are [$\alpha$/Fe]=0.3 (see Section\,\ref{alpha}) SSP model grids from S07 (dashed lines, [Fe/H]=-1.3 -0.7, -0.4, 0.0, 0.2) and T11 (solid lines, [Fe/H]=-1.35 -0.33,0.0, 0.35, 0.67). 
We have defined [MgFe] according to \cite{gonzalez93}, where [MgFe]=(Mg$b<$Fe$>$)$^{1/2}$ and $<$Fe$>$=(Fe 5270 + Fe 5335)/2.
The model grid ages are indicated in panel (a). 

In the upper panels (a, b and c) of Fig.\,\ref{indices_plot} we show the diagnostic diagrams for the newly obtained FORS2 sample.
In these panels, the big upside-down red triangle represents the median value for the GCs of the FORS2 sample.
The different colours and symbols indicate the GCs in common with the LRIS samples of L03, B05 and the new ones obtained with FORS2 according to the legend in panel (a).
The middle (d, e and f) and bottom (g, h and i) panels show the diagnostic diagrams for the B05 (indicated by green squares) and L03 (indicated by orange circles) samples respectively.
In the middle panels, the big red squares represent the median value for the GCs of the B05 sample. The black squares in panel (d) represent the medians of the three populations defined by B05. In the bottom panels the big red circles represent the median value for the GCs of the L03 sample.

The FORS2 sample confirms the metallicity spread from L03 and B05. The [MgFe] values fall between the tracks of [Fe/H] $\sim-1.3$ to $0$ for S07 models and from $\sim-1.35$ to 0.35 for T11 models. 
The mean metallicity of metal-poor GCs in this galaxy is estimated to be $-1.63$ by \cite{bs06} using the \cite{peng06} empirical colour-metallicity relation. Being our lower FORS2 metallicity value slightly higher than this, we are not sampling the lowest metallicity GCs. This is because many of the bluest GCs are quite faint so they are under-represented in this sample. The lack of blue GCs in the FORS2 sample can also be seen in Fig.\,\ref{LIRISmatchFORS2}. 

Overall in the three different kinds of diagnostic diagrams, the FORS2 measurements are distributed in a similar way to the measurements of B05 and very differently from L03.
The new data (and B05) do not suggest any tendency for the metal-rich clusters to be younger. This is, instead, what the L03 sample seems to suggest.
Most objects in the new sample have ages consistent with $8$\,Gyrs or more according to T11 in the three different diagnostic diagrams, similarly to B05.
This is not true for the L03 sample where more objects are consistent with ages $\la8$\,Gyrs. Age estimates for the L03 sample deviate the most from FORS2/B05 when one uses the H$\delta_A$ - [MgFe] diagnostic diagram. In such plot most objects have ages consistent with models $\la5$\,Gyrs.
Note that, due to the shape of the two different model grids, regardless of the diagnostics diagram, if one compares a metal-poor GC to the S07 model grid, the age inferred will be slightly younger than if it was inferred using the T11 model grid.  
On the other hand if one compares a metal-rich GC to the S07 model grid the age inferred will be older than if it was inferred using the T11 models.

L03 showed a curious correlation between H$\beta$ and metal sensitive indices such as [MgFe] and $<$Fe$>$: the larger the H$\beta$, the larger [MgFe] and $<$Fe$>$. This can be seen in panel (g). This suggested that the majority of the metal-rich clusters in that sample were younger than the metal-poor objects. As in B05, this is not seen in the FORS2 sample. 
A notable feature of the metal sensitive \textit{vs.} age sensitive plots for the B05 sample was the grouping of objects into three distinct areas. This was suggested by B05 as consistent with the presence of three old sub-populations. In panel (d) median values of these three populations are shown as filled black squares. This division is not clearly seen for the FORS2 sample. This is probably due to the fact that our sample does not span the full metallicity range while the B05 metallicity range is more complete and representative of the GC system of NGC\,4365. 

The blue horizontal branch (HB) is a hot component of an old stellar population that has roughly the same effect as a younger population in some of the Balmer indices.
Since it essentially contributes at short wavelengths, the relative contribution of the blue HB to H$\delta_{A}$ is larger than it is for H$\beta$. Hence, ages inferred from H$\delta_{A}$ for blue HB GCs should be younger than the ones derived through H$\beta$. For Red HB GCs the ages derived through these indices should be roughly the same (see e.g. \citealt{schiavon04})
One might think that part of the differences between the FORS2/B05 and the L03 sample might be due to differences between the horizontal branch morphologies of the samples. 
The average age for the L03 sample inferred from H$\delta_{A}$ is younger than the one inferred from H$\beta$, regardless of the model.
For instance, following T11 models the median age for L03 according to H$\delta_A$ (H$\beta$) is consistent with $\sim3$\,Gyrs ($\sim8$\,Gyrs). 
As already mentioned, the ages inferred for FORS2 and B05 H$\delta_{A}$ and H$\beta$ are consistent with $\ga8$\,Gyrs.
Therefore at first sight one might indeed conclude that the L03 sample has on average more blue HB GCs than the B05 and FORS2 samples.
However, from the uncertainties on the individual data points it is difficult to give any definitive answer. 

The GCs that are in common with all samples are shown as magenta stars in all panels of Fig.\,\ref{indices_plot} and are indicated by their ids of Table\,\ref{GCs} in panels (a), (d) and (g). 
Two of the three GCs that are in common to the three samples are found to have very old ages in this study.
For the FORS2 sample, objects 8 and 13 are compatible with models $>14$\,Gyrs (for both sets of models), whereas object 9 with models $3<$age$<5$\,Gyrs (for both sets of models) on the [MgFe] \textit{vs.} H$\beta$ plot.
In L03 GCs 8 and 9 were found to be of intermediate-age, being consistent with models $<3$\,Gyrs in this same diagram while object 13 is consistent with old models (e.g. T11, $\sim14$ Gyrs)  
For B05, objects 8 (and 9) are consistent with ages $>$14\,Gyrs ($>$11\,Gyrs) following T11 models while object 13 is consistent with the 8\,Gyr model.
While it is hard to infer the age of object 9 based on the three samples, the most extreme outlier in L03, object 8 appears old in the current and in the B05 samples.  

In summary, there are differences between inferred ages and metallicities of all datasets but overall FORS2 measurements tend to agree better with the ones of B05 than with L03. FORS2 derived ages are consistent with the range in ages 8-14\,Gyrs, with significant errors and uncertainties on the model.

\begin{table*}
\begin{scriptsize}
\begin{center}
 \caption{The FORS2 Lick/IDS indices of the GCs of NGC\,4365 used in this paper.}
 \label{lickindices}
 \begin{tabular}{cccccccccccccc}
  \hline
ID  &           H$\beta$ (\AA)                & H$\gamma_A$ (\AA)     &    H$\delta_A$ (\AA)               &     Fe5270 (\AA)  &      Fe5335 (\AA)            &         $Mg\,b$ (\AA)        &   CN${_1}$ (mag)            &       CN${_2}$ (mag)                       &              Mg${_2}$(mag)                          \\
\hline                                                                                                                                                                                                                                                                     
07  &      1.29  $\pm$  0.14    &        -3.76    $\pm$   0.26       &    0.12  $\pm$  0.28     &          2.06   $\pm$ 0.18  &     1.94  $\pm$ 0.21      &         2.85   $\pm$    0.15   &  0.05     $\pm$  0.01     &       0.11    $\pm$      0.01         &          0.16        $\pm$         0.01               \\
08  &       1.33  $\pm$  0.23    &       -3.36   $\pm$   0.41       &    -0.59 $\pm$  0.41     &           1.97  $\pm$ 0.27  &      2.35 $\pm$ 0.34      &          2.32  $\pm$    0.26   &   0.07    $\pm$   0.01    &        0.09   $\pm$      0.01         &           0.18       $\pm$          0.01            \\
09  &      2.41   $\pm$  0.13    &       -2.79   $\pm$   0.24       &    -0.82 $\pm$  0.24     &          2.38   $\pm$ 0.15  &     1.67  $\pm$ 0.17     &         2.89   $\pm$    0.16   &    0.09   $\pm$    0.01   &        0.13   $\pm$      0.01         &            0.18       $\pm$           0.01            \\
10  &       2.18  $\pm$  0.09    &        -0.81  $\pm$   0.15       &    2.36  $\pm$  0.17     &           1.55  $\pm$ 0.11  &      1.09 $\pm$ 0.12     &          1.71  $\pm$    0.10   &    -0.06  $\pm$    0.01   &        -0.01  $\pm$      0.01         &            0.09     $\pm$           0.01           \\
11  &       2.27  $\pm$  0.15   &       -3.67   $\pm$   0.25       &    -0.56 $\pm$  0.27     &           2.32  $\pm$ 0.17  &      1.85  $\pm$ 0.17     &         2.72   $\pm$   0.15   &      0.02  $\pm$    0.01    &        0.06   $\pm$     0.01          &           0.13      $\pm$          0.01             \\
12  &       1.74  $\pm$  0.17    &       -3.96  $\pm$   0.32       &    -1.11 $\pm$  0.36     &           2.57  $\pm$ 0.19  &      1.81 $\pm$ 0.24      &          3.71  $\pm$    0.18   &     0.05  $\pm$    0.01   &        0.09   $\pm$      0.01         &            0.19     $\pm$           0.01            \\
13  &      1.28   $\pm$  0.24    &        0.96   $\pm$   0.37       &    1.43  $\pm$  0.39     &           2.38  $\pm$ 0.30  &      1.18 $\pm$ 0.37      &          1.64  $\pm$    0.26   &   -0.06   $\pm$   0.01    &        -0.03  $\pm$      0.01         &           0.06      $\pm$          0.01             \\
14  &     2.00    $\pm$  0.16    &      -1.06    $\pm$   0.28       &    0.20  $\pm$  0.28     &         1.71    $\pm$ 0.18  &    1.63   $\pm$ 0.23      &        3.28    $\pm$    0.17   &   0.09    $\pm$   0.01    &       0.14    $\pm$      0.01         &           0.13        $\pm$          0.01            \\
15  &     2.19    $\pm$  0.19    &      -2.15    $\pm$   0.33       &    0.78  $\pm$  0.33     &         2.38    $\pm$ 0.22  &    1.73   $\pm$ 0.26      &        2.24    $\pm$    0.22   &   0.06    $\pm$   0.01    &       0.09    $\pm$      0.01         &           0.16        $\pm$          0.01            \\
16  &       2.57  $\pm$  0.15    &        -2.07  $\pm$   0.23       &    2.01  $\pm$  0.26    &           2.22  $\pm$ 0.16  &      0.84 $\pm$ 0.21      &          2.33  $\pm$    0.16  &     -0.05  $\pm$    0.01   &       -0.02    $\pm$      0.01          &          0.14    $\pm$          0.01              \\
17  &       2.18  $\pm$  0.14    &       -3.73   $\pm$   0.24       &    -1.16 $\pm$  0.29     &           2.04  $\pm$ 0.17  &      2.34 $\pm$ 0.19      &          3.41  $\pm$    0.15   &    0.08   $\pm$   0.01    &        0.12   $\pm$      0.01         &           0.19      $\pm$          0.01            \\
18  &       1.93  $\pm$  0.14    &       -3.37   $\pm$   0.26       &    -1.21 $\pm$  0.26     &           1.81  $\pm$ 0.16  &      1.61 $\pm$ 0.19      &          2.56  $\pm$    0.16   &    0.10   $\pm$   0.01    &        0.14   $\pm$      0.01         &           0.14      $\pm$          0.01            \\
19  &      1.76   $\pm$  0.11    &      -3.54    $\pm$   0.19       &    0.40  $\pm$  0.20     &          1.92   $\pm$ 0.13  &     1.71  $\pm$ 0.14      &         2.91   $\pm$    0.13   &   0.04    $\pm$   0.01    &       0.09    $\pm$      0.01         &           0.15       $\pm$          0.01            \\
20  &     1.86    $\pm$  0.14    &      -1.53    $\pm$   0.20       &    1.11  $\pm$  0.23     &         1.52    $\pm$ 0.15  &    1.41   $\pm$ 0.16      &        2.02    $\pm$    0.14   &   0.01    $\pm$   0.01    &       0.06    $\pm$      0.01         &           0.10        $\pm$          0.01            \\
21  &      1.68   $\pm$  0.24    &       -0.68   $\pm$   0.40       &    -0.29 $\pm$  0.43     &          1.49   $\pm$ 0.27  &     1.49  $\pm$ 0.35      &         1.92   $\pm$    0.26   &    0.05   $\pm$   0.01    &        0.05   $\pm$      0.01         &           0.10       $\pm$          0.01            \\
22  &    2.18     $\pm$  0.17    &     -1.18     $\pm$   0.28       &   1.53   $\pm$  0.28     &        1.11     $\pm$ 0.18  &   0.84    $\pm$ 0.23      &       1.91     $\pm$    0.22   &   0.01    $\pm$   0.01   &      0.03     $\pm$      0.01         &            0.10       $\pm$           0.01           \\
23  &      1.59   $\pm$  0.13    &      -4.70    $\pm$   0.26       &    0.52  $\pm$  0.26     &          2.13   $\pm$ 0.14  &     1.90  $\pm$ 0.17      &         5.00   $\pm$    0.13   &   0.06    $\pm$   0.01    &       0.10    $\pm$      0.01         &           0.24       $\pm$          0.01            \\
24  &       1.61  $\pm$  0.18    &       -3.67   $\pm$   0.35       &   -1.01  $\pm$  0.35     &           2.32  $\pm$ 0.22  &      1.56 $\pm$ 0.25     &          3.35  $\pm$    0.21   &     0.08  $\pm$    0.01   &        0.10   $\pm$      0.01         &            0.19     $\pm$           0.01           \\
25  &       2.51  $\pm$  0.15    &        -0.76  $\pm$   0.24       &    1.71  $\pm$  0.24     &           1.53  $\pm$ 0.18  &      0.90 $\pm$ 0.20      &          1.75  $\pm$    0.17   &   -0.04   $\pm$   0.01    &        -0.01  $\pm$      0.01         &           0.09      $\pm$          0.01            \\
27  &        2.02  $\pm$  0.20   &         -0.04   $\pm$   0.32     &    0.00 $\pm$  0.36      &        1.93  $\pm$ 0.24     &    1.66 $\pm$  0.26        &         1.55  $\pm$    0.25   &    0.05   $\pm$   0.01     &     0.09     $\pm$      0.01          &          0.09      $\pm$          0.01          \\
28  &     2.07    $\pm$  0.099   &      -1.44    $\pm$   0.17       &    1.23  $\pm$  0.17     &         2.03   $\pm$ 0.11  &      1.68   $\pm$  0.13      &        2.26  $\pm$    0.11   &    0.01   $\pm$   0.01    &       0.05   $\pm$        0.01          &         0.12    $\pm$           0.01             \\
     
 \hline
 \end{tabular}
 \end{center}
 \end{scriptsize}
 \end{table*}

\subsection{[$\alpha$/Fe] abundance ratios}
\label{alpha}
\begin{figure}
\begin{center}
\includegraphics[width=7cm, angle=90]{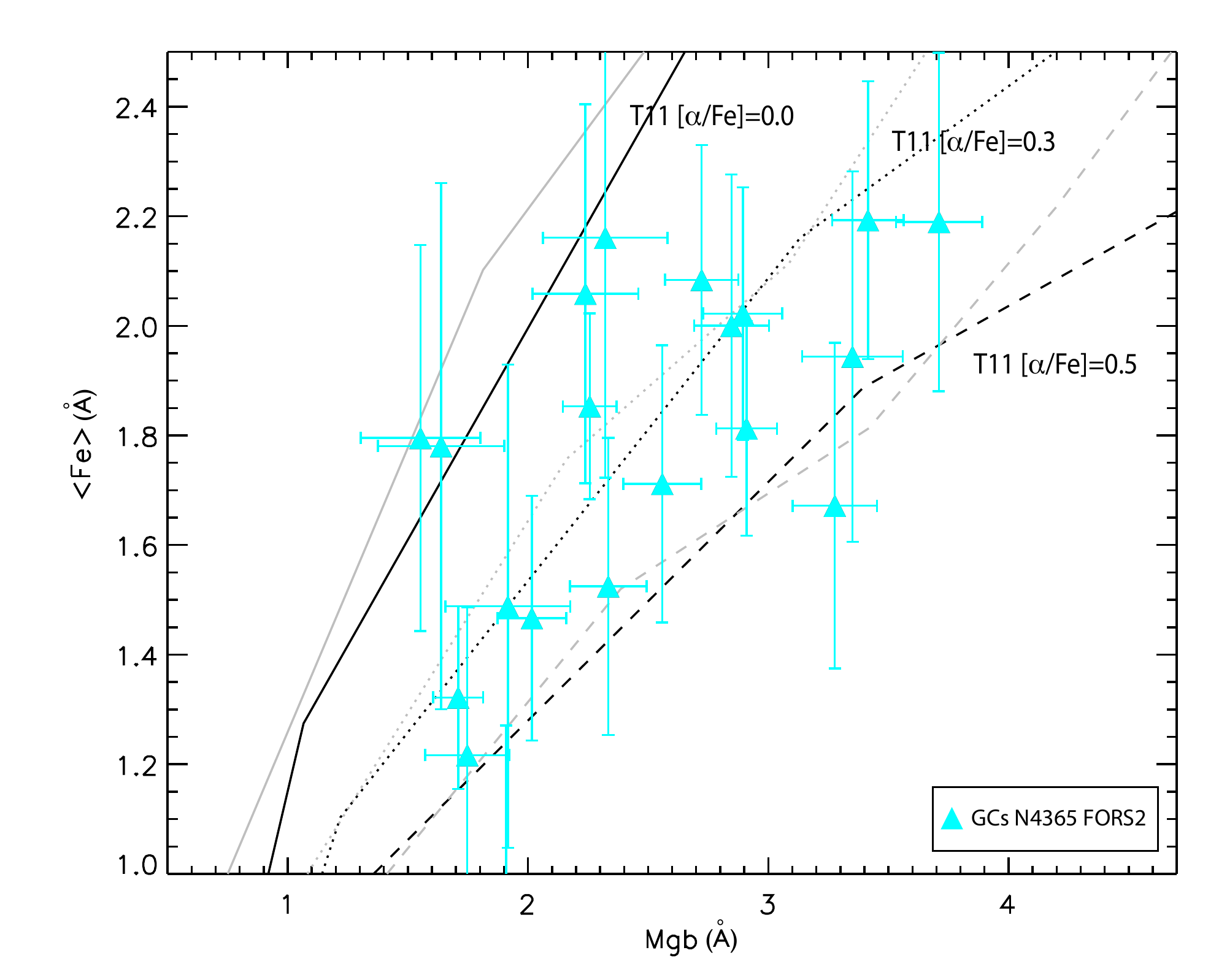}
\caption{Mg$b$ \textit{vs.} $<$Fe$>$ diagram for the FORS2 sample of GCs in NGC\,4365. Symbols are as in Fig. \ref{indices_plot}. Overlaid are T11 SSP models for 3\,Gyrs (in grey) and 14\,Gyrs (in black) and different [$\alpha$/Fe] values as indicated. The S07 models are very close to the T11 ones in this diagram.}
\label{mgb_fe_plot}
\end{center}
\end{figure}

The [$\alpha$/Fe] ratio is largely used to constrain the formation time scales of stellar populations in galaxies.
The $\alpha$-elements (e.g. O, Mg, Ca, Si, Ti) are released early in the evolution of a galaxy by Type II SNe (a core-collapse star, result of the evolution of a single massive star). The Fe-peak elements are mainly produced by type Ia SNe (result of a binary interaction between a white dwarf and a giant star or alternatively it is the result of a double degenerate white dwarf - white dwarf merger) on longer time scales. 
In this section we investigate the [$\alpha$/Fe] values for the GCs in NGC\,4365.
In Fig.\,\ref{mgb_fe_plot} we compare the values of Mg$b$ (a proxy for $\alpha$-elements.) and $<$Fe$>$ of the GCs with different [$\alpha$/Fe] models. T11 SSP models are shown for the ages of 3\,Gyrs and 14\,Gyrs and [$\alpha$/Fe]=0, 0.3 and 0.5. The models for the different ages and same [$\alpha$/Fe] values are very close together, showing that this comparison is only weakly dependent on age. S07 models fall very close to the T11 ones in such a diagram.

The [$\alpha$/Fe]=0.3 T11 models are a good match to the data. However, as found in B05, some of the GCs have supersolar [$\alpha$/Fe]. 
In a study of GCs in several early-type galaxies, \cite{puzia05} finds a mean [$\alpha$/Fe] value of $\sim$0.5. 
There is much scatter in Fig.\,\ref{mgb_fe_plot} and a few GCs appear to have slightly lower [$\alpha$/Fe], between 0 and 0.1 when looking at their absolute values.
However, within the uncertainties most objects are consistent with the [$\alpha$/Fe]=0.3 model.

\subsection{C \& N sensitive indices}
In addition to $\alpha$-elements, GCs show a variety of abundance anomalies with respect to the solar mixture (see \citealt{gratton04} for a review of this issue on Milky Way GCs). The most notable anomaly in extragalactic GCs is the CN-enhancement (\citealt{bs06}). This anomaly has been reported for the Milky Way, M31 (\citealt{burstein84}; \citealt{bh91}), Fornax dSph, NGC 3115, NGC\,3610 (\citealt{trager04} and references therein) and NGC\,1407 (\citealt{cenarro07}).
CN enhancement seems an almost generic property of (extragalactic) GCs except for LMC GCs (\citealt{trager04}). This enhancement appears to be due to an excess of N over the solar mixture (\citealt{burstein04}). 
The fact that CN enhancement is being found in several GC systems implies that very high N abundances are a generic feature of the early chemical evolution of GC systems (\citealt{bs06}).
In the study of NGC\,1407 by \cite{cenarro07} it was noted that compared to metal-poor GCs, the metal-rich GCs exhibit a striking N overabundance. This behaviour was found to be consistent with the outcome of the CNO nucleosynthesis cycle. For low and intermediate mass stars C yields decrease with increasing initial metallicity and N yields slightly increase with metallicity (\citealt{chiappini03}). N is formed by consumption of C and O already present in the stars. 

More recently however, \cite{schiavon12} presented evidence that the CN differences between M31 and Galactic clusters are mostly due to data calibration uncertainties. They suggest that the chemical compositions of Milky Way and M31 globular clusters are not substantially different, and that there is no need to invoke enhanced nitrogen abundances to account for the optical spectra of M31 globular clusters. Nevertheless, it is worthwhile to investigate whether there are any differences in CN in our FORS2 measurements of GCs in NGC\,4365 and other GC systems.

\begin{figure}
\begin{center}
\includegraphics[width=8cm]{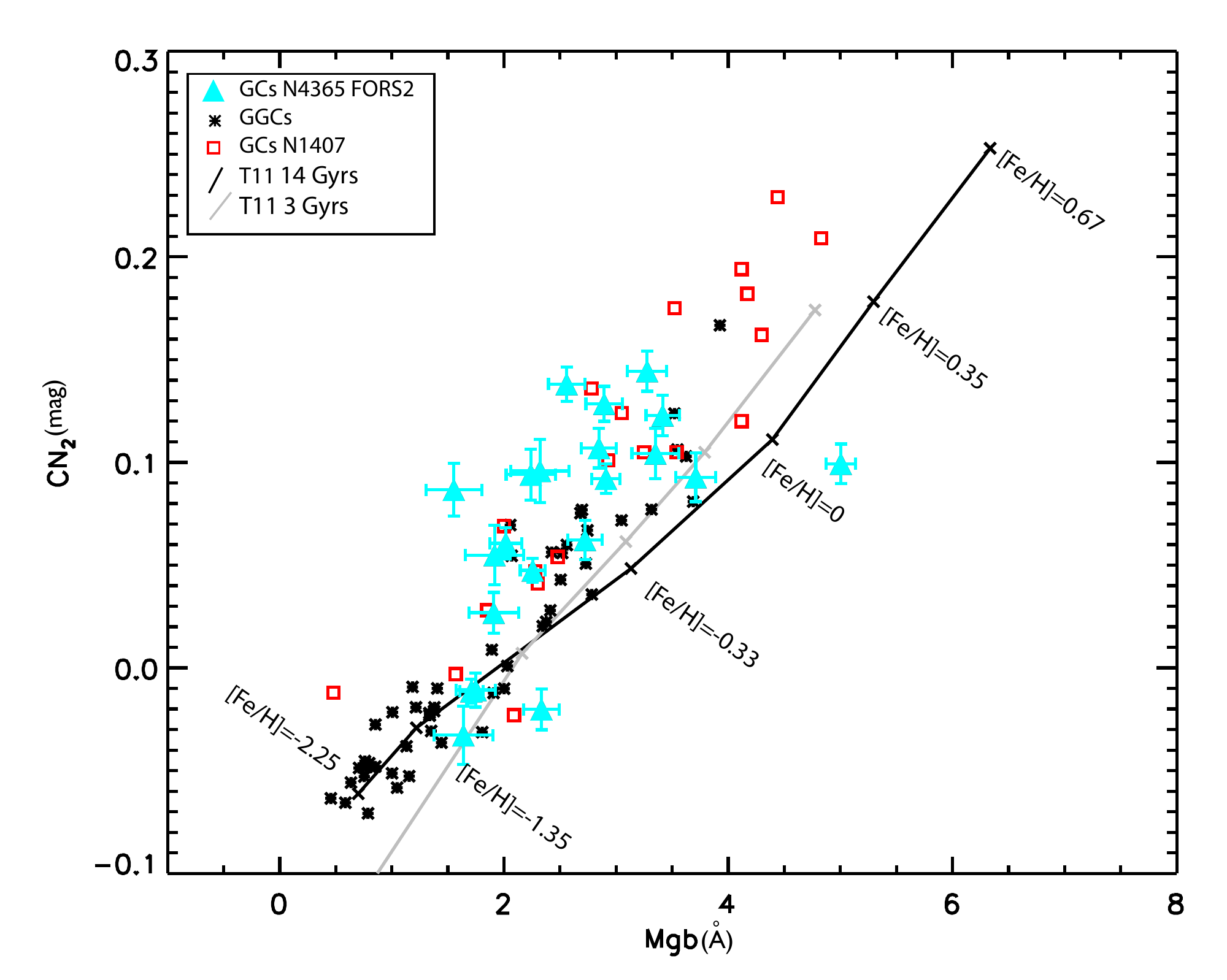}
\caption{Mgb \textit{vs.} CN$_{2}$ for the GCs in NGC\,4365 GCs according to the legend. Asterisks are Galactic GCs from \protect \cite{schiavon05} and \protect \cite{puzia02b} and open red squares are GCs from NGC\,1407 (\protect \citealt{cenarro07}). T11 SSP models for [$\alpha$/Fe]=0.3 and 2 different ages are overlaid as indicated in the legend. The different metallicity values are indicated for the 14Gyr old model (black line).}
\label{CN_mgb} 
\end{center}
\end{figure}

In Fig.\,\ref{CN_mgb} the Mgb \textit{vs.} CN$_{2}$ diagnostic diagram is shown for the GCs in NGC\,4365 (FORS2 sample), NGC\,1407 (\citealt{cenarro07}) and Milky Way GCs from \cite{puzia02b} and \cite{schiavon05}. 
T11 ([$\alpha$/Fe]=0.3) 3 and 14\,Gyr SSP models are overplotted as indicated in the legend. 
The more metal-rich clusters present CN enhancement at fixed Mgb strength, as found for other galaxies as noted before. This enhancement is larger than for the Milky Way GCs.
The NGC\,4365 data is in close agreement to that of NGC\,1407 in Fig.\,\ref{CN_mgb}. However, our sample of GCs does not go to very metal-rich values as the one of NGC\,1407.
Therefore we do not find the clear trend for the very metal-rich GCs to be strikingly CN enhanced as did \cite{cenarro07} for NGC\,1407.  
One object deviates substantially from the group, this is object 23. It has a relatively normal CN$_2$ value compared to the rest of the sample but a high Mgb value, $\sim5$\AA.
Interestingly, this object with $<$Fe$>$$\sim$\,2\AA~and [Mgb]$\sim$\,3.5\AA~is also an outlier in panel (c) of Fig.\,\ref{indices_plot}, having quite a high H$\delta_A$ while low $H\beta$ and $H\gamma_A$ for its high [MgFe] value. Its high H$\delta_A$ and low $H\beta$ (and $H\gamma_A$) suggests that this object could be a blue horizontal branch cluster (see discussion in Sect. 3.1). It could also be that its Mgb value is overestimated giving it an artificially high metallicity.

\subsection{Comparison with photometry}
\begin{figure}
\begin{center}
\includegraphics[width=5cm]{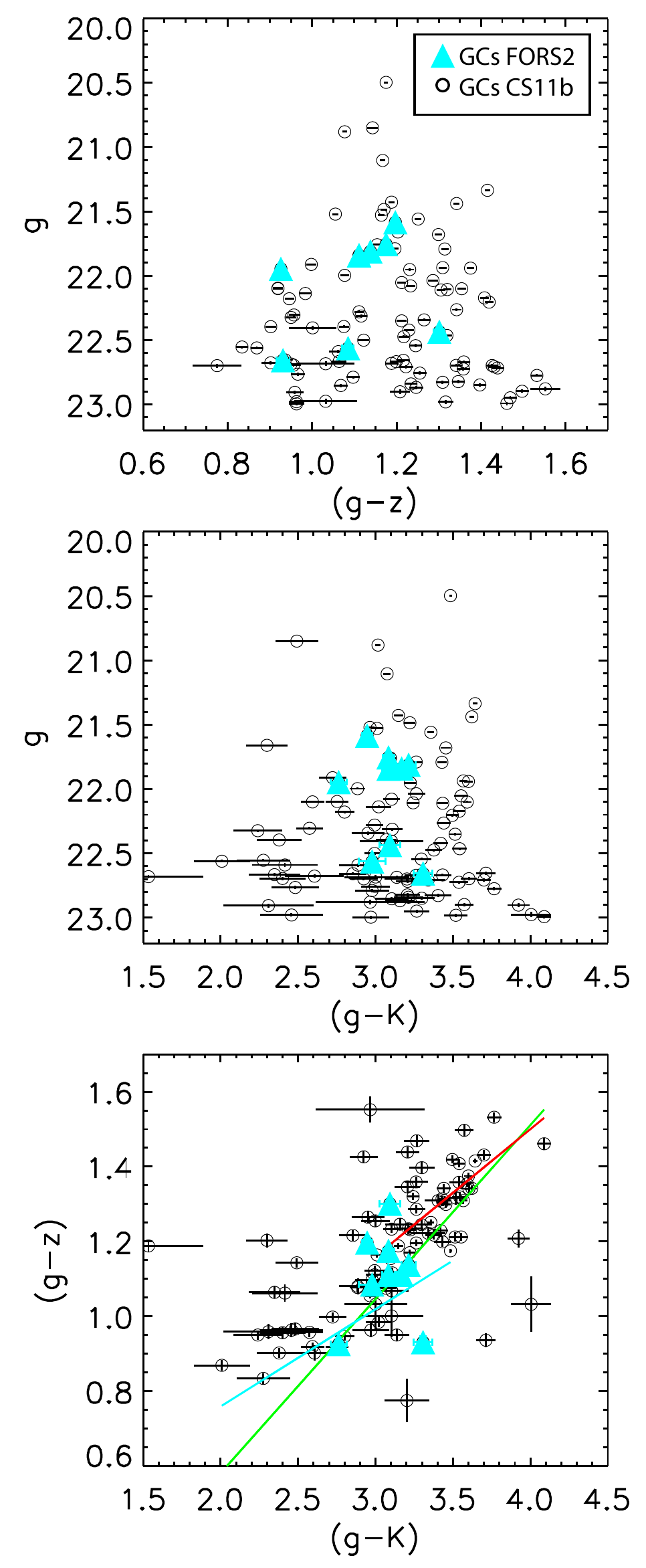}
\caption{Comparison of optical/near-infrared photometry and the FORS2 spectra, GCs with spectroscopic measurements are marked according to the legend. CMDs (\textit{upper and middle panels)}. Colour-colour diagram $(g-K)$ \textit{vs.} $(g-z)$ (\textit{bottom panel)}. The green line is an empirical best-fit relation for the joint GC system of NGC\,4486 and NGC\,4649 (\citealt{paper2}). Similarly, the blue and red lines are empirical best-fit relations when separating in blue and red clusters.}
\label{LIRISmatchFORS2}
\end{center}
\end{figure}

Of the 22 spectra obtained, a match to 9 objects was found in the LIRIS/ACS optical/near-infrared data set of \cite{paper1}. 
In Fig. \ref{LIRISmatchFORS2} colour-magnitude diagrams (CMD) and the $(g-K)$ \textit{vs.} $(g-z)$ colour-colour diagram are shown with GCs with spectroscopic measurements with symbols according to the legend.
Note that while for the $(g-z)$ \textit{vs.} $g$ CMD the objects with spectra span the blue to red colour ranges, in the $(g-K)$ \textit{vs.} $g$ they are much more concentrated towards intermediate $(g-K)$ values. 
In the colour-colour diagram this is clearly seen. 
This is a coincidence due to the photometric scatter. Blue clusters are generally faint in $K$ and therefore the photometric errors are larger in $(g-K)$ broadening the CMD in the middle panel to the blue. However the photometric errors of our GCs with spectroscopic measurements are very small and are not spread to the blue in the same diagram.
There is one object, 13, which has a $\delta$ value consistent with a younger value than the joint GC system of NGC\,4486 and NGC\,4649 (\citealt{paper2}). In the optical/near-infrared photometric study of \cite{paper2} the position of the $\delta$ parameter in the $(g-K)$ \textit{vs.} $(g-z)$ diagram was used to define a relative age (to the joint GC system of NGC\,4486 and NGC\,4649).
However, following H$\beta$ (H$\delta_A$) (see Fig.\,\ref{indices_plot}) object 13 has an inferred T11 age $\sim$14\,Gyr ($\sim$8\,Gyr) according to the FORS2 measurements. Such object could be thus a blue HB GC according to the FORS2 measurements but not according to B05 measurements (see discussion in Sect.\,\ref{balmer}). In the L03 sample, object 13 has a large uncertainty and within the errors it is consistent with the 5\,Gyr T11 model for the three different Balmer line indices.

\begin{figure}
\begin{center}
\includegraphics[width=5cm]{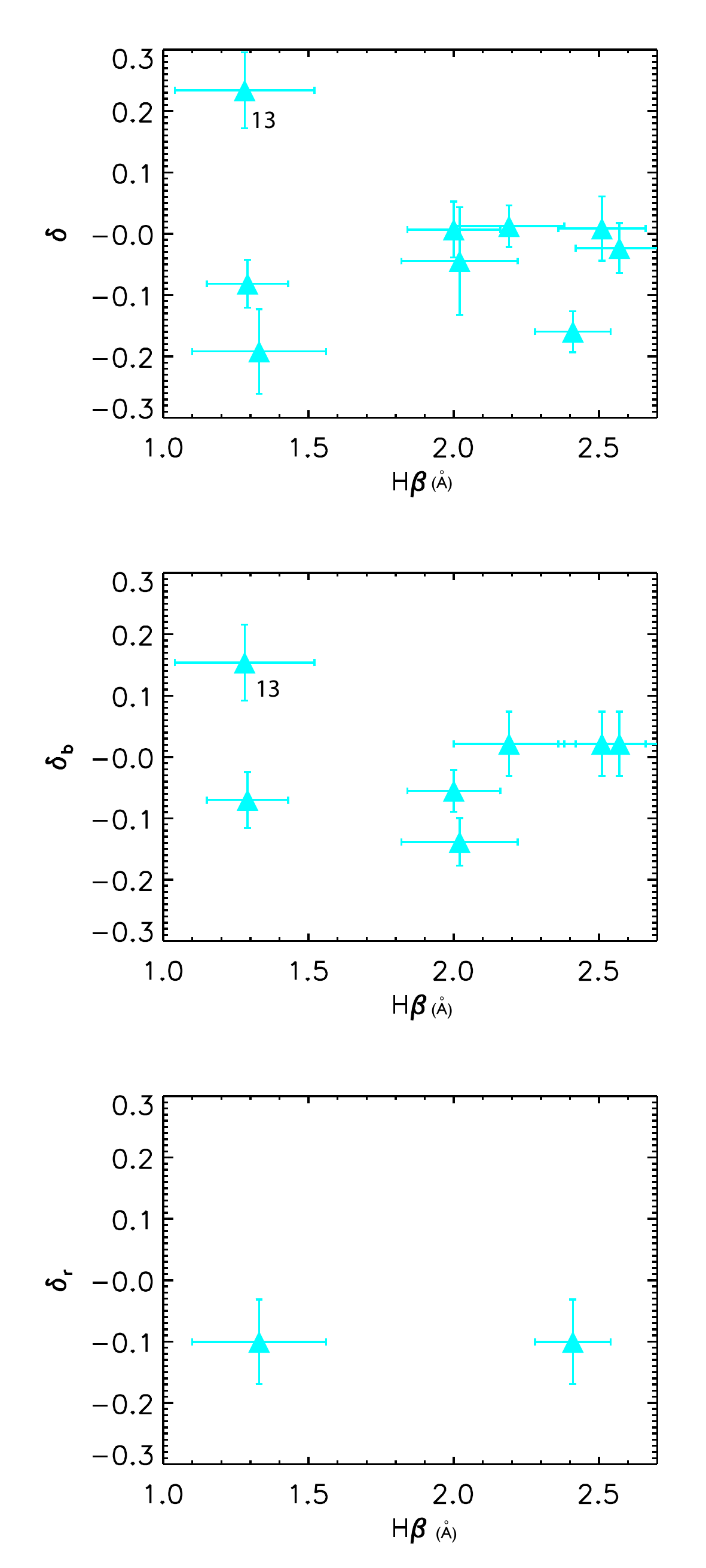}
\caption{H$\beta$ \textit{vs.} the $\delta$, $\delta_{blue}$ and $\delta_{red}$ parameters from \protect \cite{paper2}, for the GCs of that study with FORS2 measurements. Object 13 has its id indicated according to its designation of Table\,1.}
\label{LIRISmatchFORS2_balmer}
\end{center}
\end{figure}

The derivation of absolute ages and metallicities is beyond the scope of this paper.
However, one can still investigate in a general sense how age and metallicity measurements relate to photometric measurements. 
In the following we use some age and metallicity sensitive indices to do so.  
Fig.\,\ref{LIRISmatchFORS2_balmer} shows a comparison between the photometrically determined $\delta$, $\delta_{blue}$ and $\delta_{red}$ from \cite{paper2} \textit{vs.} the H$\beta$ index. 
There is no obvious relation between the different $\delta$ parameters and the H$\beta$ index.  
Coincidentally most objects appear to have very similar $\delta$ values; nearly equal or below 0 (except for object 13 which is indicated in the Fig., see discussion above). However their H$\beta$ values vary a lot, ranging from 1.3 to 2.5. H$\beta$ depends not only on age but also on metallicity (see Fig.\,\ref{indices_plot}).
The variations seem larger than what one would expect from Poisson errors alone. This is not uncommon for this kind of study (e.g. B05) and probably depends on the exact mix of stars in the clusters. 
For instance, \cite{schiavon02} find that an error of $\pm$\,0.1 dex in the luminosity function of giants brighter than the HB translates into an error
of $\pm$\,1\,Gyr in the spectroscopic age inferred from Balmer lines.
At this point it is unclear whether the lack of a relation between the $\delta$ parameter and H$\beta$ is real or if it is an effect of a small number of overlapping sources between this study and that of \cite{paper2} combined with large H$\beta$ variations.

\begin{figure}
\begin{center}
\includegraphics[width=9cm]{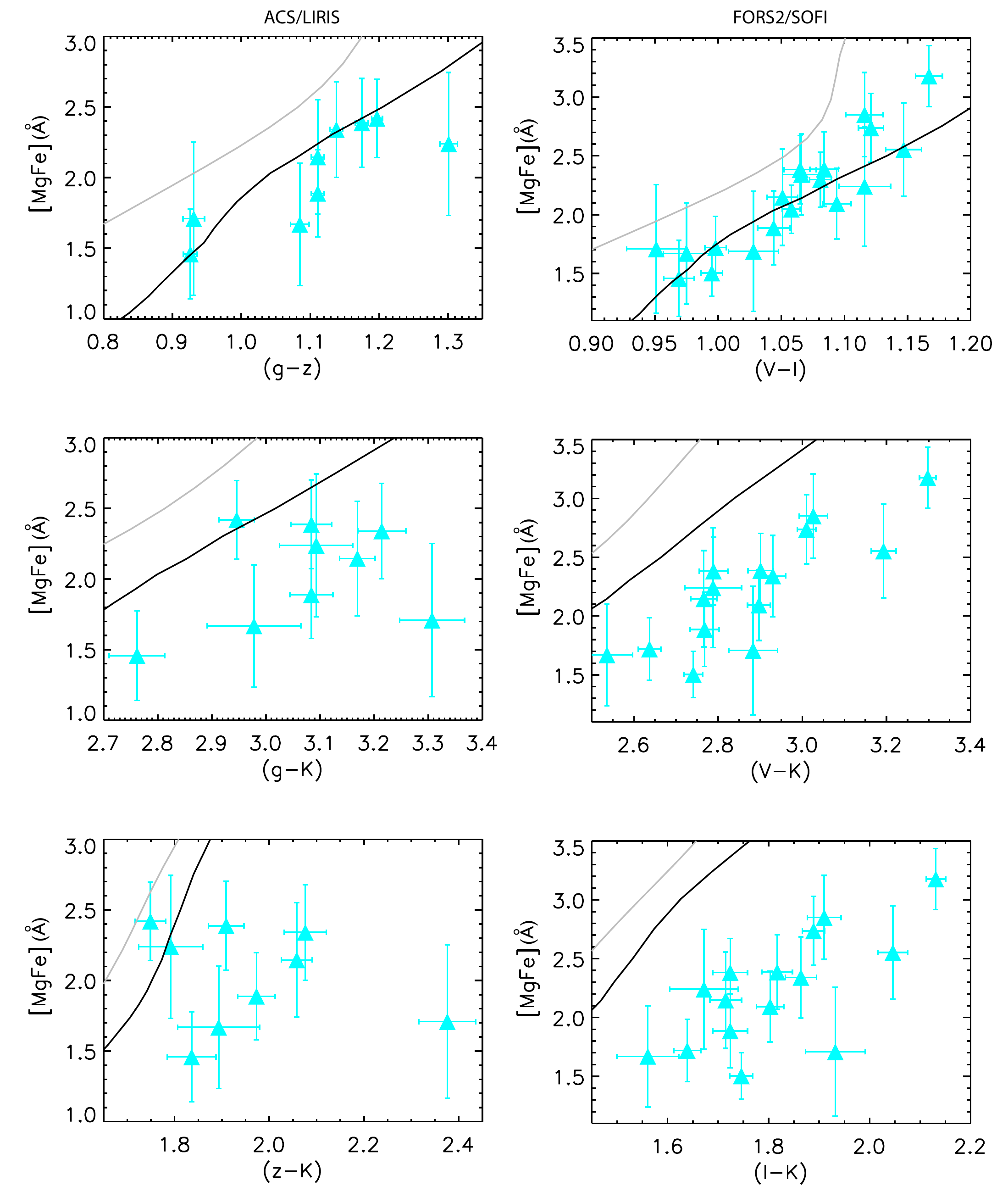}
\caption{Different colours \textit{vs.} the [MgFe] index. The $(g-z)$, $(g-K)$ and $(z-K)$ are ACS/LIRIS from \protect \cite{paper2} and $(V-I)$, $(V-K)$ and $(I-K)$ are FORS2/SOFI from \protect \cite{lbs05}. Overlaid are YEPS models with HB for 3 (in gray) and 14\,Gyrs (in black). Models without HB are very close to models with HB.}
\label{LIRIS_SOFImatchFORS2_mgfe}
\end{center}
\end{figure}

It is of great interest to study the relation between metalicity and colours due to its implications on the colour bimodality of GC systems (e.g. \citealt{yoon06}; \citealt{cb07}; \citealt{yoon11a}; \citealt{yoon11b}; \citealt{paper3}; \citealt{blakeslee12}).
The left panels of Fig.\,\ref{LIRIS_SOFImatchFORS2_mgfe} show the $(g-z)$, $(g-K)$ and $(z-K)$ from \cite{paper2} \textit{vs.} the [MgFe] index.
Besides a few outliers, the $(g-z)$ and $(g-K)$ colours correlates reasonably well with the metal sensitive index [MgFe]. However, there is more scatter in the $(g-K)$ plot and even more in the $(z-K)$ one.
The right panels of Fig.\,\ref{LIRIS_SOFImatchFORS2_mgfe} show the $(V-I)$, $(V-K)$ and $(I-K)$ colours from \cite{lbs05} \textit{vs.} the [MgFe] index. The full sample of FORS2 objects has $V$ and $I$ photometry and only 4 do not have $K$. 
For the colours from \cite{lbs05} there is a good correlation with the metal sensitive index [MgFe]. As in the left panels of Fig.\,\ref{LIRIS_SOFImatchFORS2_mgfe}, there is more scatter for the optical/NIR colours if compared to the optical colour alone.

YEPS\footnote{YEPS models can be downloaded from http://web.yonsei.ac.kr/cosmic/data/YEPS.htm}
models with HB variation are shown for 2 different ages in the different panels of Fig.\,\ref{LIRIS_SOFImatchFORS2_mgfe}.
Models with the HB variation are very close to models without this variation in such a diagram.
In the top panels it is clear that the 14\,Gyr old model is a better match to the data than the 3\,Gyr one.
The models for optical/NIR colours and the [MgFe] index are not a good match to the data. 
Interestingly, such models are too blue by $\sim$0.2\,mag, for instance in $(g-K)$. This is similar to what was found by \cite{paper2}. In that study, (see their Fig. 2) widely used SSP models were offset from the data by the same value when optical/NIR colours were used.
This is probably because of the AGB modeling (see discussion in \citealt{paper2}).
However, note that in Fig.\,\ref{LIRIS_SOFImatchFORS2_mgfe} the younger models fall even further away from the data. This is the reverse of what we see in \cite{paper2}. Therefore in this case, the offset cannot be explained by younger ages. 
Similarly, \cite{lbs05} compared metallicity measurements of L03 and B05 with their $(V-I)$ and $(V-K)$ photometry and found that the colour-metallicity relation for $(V-K)$ was offset from the data.

\section{Summary \& Conclusions}
We have revisited the GC system of NGC\,4365 using VLT/FORS2 high S/N spectroscopy.
Below we summarize our principal findings: 
\begin{enumerate}
\item Based on Balmer line indices the FORS2 GCs have ages consistent with the age range 8\,Gyrs - 14\,Gyrs.  
These ages are in accordance with the study of \cite{brodie05} and deviate from the conclusions of \cite{larsen03}. 
\item The observed FORS2 GCs exhibit a metallicity spread, [Fe/H] between $\sim-1.3$ and $\sim0.3$ according to T11 models. 
The majority of them have [$\alpha$/Fe] values consistent with $\sim$\,0.3.
Also, the majority of the observed GCs in NGC\,4365 appears to be CN enhanced if compared to Milky Way GCs. However if compared to the GC system of NGC\,1407 this enhancement is less strong. However, this latter sample spans a broader metallicity range, probing higher metallicities than ours for NGC\,4365. 
\item A comparison between metal sensitive indices and photometry from \cite{paper2} and \cite{lbs05} is performed. While the match between our FORS2 measurements and that of \cite{paper2} is small, the \cite{lbs05} photometric sample and our FORS2 measurements have a significant overlap. We find that $(V-I)$, $(V-K)$ and $(I-K)$ colours correlate with the metal sensitive index [MgFe]. However, the scatter is too large to draw any conclusions regarding the shape of the colour-metallicity relation.
We find that old ($\sim14$\,Gyrs) YEPS SSP models for optical colours are a good match to the data. 
However, we also find that these SSP models for optical/NIR colours are too blue by $\sim0.2$\,mag to be consistent with the data, similarly to what has been seen in \cite{paper2} for other widely used models. This offset is not consistent with an age effect as it is even larger for younger models.
\end{enumerate}

\section*{Acknowledgments}
Ricardo Schiavon, Jonas Johansson, Jean Brodie, Duncan Forbes, Robert Proctor, Rog\'erio Riffel and Eveline Helder are thanked for useful discussions.
Based on observations collected at the European Organization for Astronomical Research in the Southern Hemisphere, Chile, under the observing program 078.B-0705(A).

\label{lastpage}
\end{document}